\documentclass[twocolumn,aps,prb,showkeys,10pt,superscriptaddress]{revtex4-1}
\usepackage{graphicx}
\usepackage{amssymb}
\usepackage{amsfonts}
\usepackage{amsmath}
\usepackage{amsbsy}
\usepackage{natbib}
\usepackage{comment}
\usepackage{color}
\usepackage[utf8]{inputenc}
\usepackage[T1]{fontenc}
\usepackage{amsthm}
\usepackage[colorlinks=true, urlcolor=blue, citecolor=blue,
linkcolor=blue]{hyperref}
\usepackage{scalerel}

\usepackage{lmodern}
\usepackage{graphicx}
\usepackage{mathtools}

\begin{document}

\title{ Time-resolved buildup of two-slit-type interference from a
single atom}

\author{Jonas W\"{a}tzel}
\affiliation{Martin-Luther-Universität Halle-Wittenberg,
Karl-Freiherr-von-Fritsch-Str. 3, 06120 Halle/Saale, Germany}
\author{Andrew  James Murray}
\affiliation{Photon Science Institute, School of Physics \&
Astronomy, University of Manchester, Manchester M13 9PL, UK}
\author{Jamal Berakdar}
\affiliation{Martin-Luther-Universität Halle-Wittenberg,
Karl-Freiherr-von-Fritsch-Str. 3, 06120 Halle/Saale, Germany}

\begin{abstract}
A photoelectron forced to pass through two atomic energy levels before receding from the residual ion
 shows interference fringes in its angular distribution as manifestation of a two-slit-type interference experiment in wave-vector space.
 This  scenario was experimentally realized by irradiating a Rubidium atom by two low-intensity continuous-wave lasers \cite{pursehousedynamic}. In a one-photon process the first laser excites the 5p  level while the second uncorrelated photon elevates the excited population to the continuum. This same continuum state can also be reached when the second laser excites  the  6p state and the first photon then triggers the ionization.  As the two lasers are weak and their relative phases uncorrelated, the coherence needed for generating the interference stems from the atom itself. Increasing the intensity or shortening the laser pulses enhances the probability that two photons from both lasers act at the same time, and hence the coherence properties of the applied lasers are expected to  affect the interference fringes.
Here, this aspect is investigated in detail, and it is shown how  tuning the temporal shapes of the laser pulses allows for tracing the time-dependence of the interference fringes. We also study the influence of applying a third laser field with a random amplitude, resulting in a random fluctuation of one of the ionization amplitudes and discuss how the interference fringes are affected.
\end{abstract}

\date{\today}
\maketitle
	
\section{Introduction}
\label{sec:1}
\begin{figure}[t]
\includegraphics[width=1.0\columnwidth]{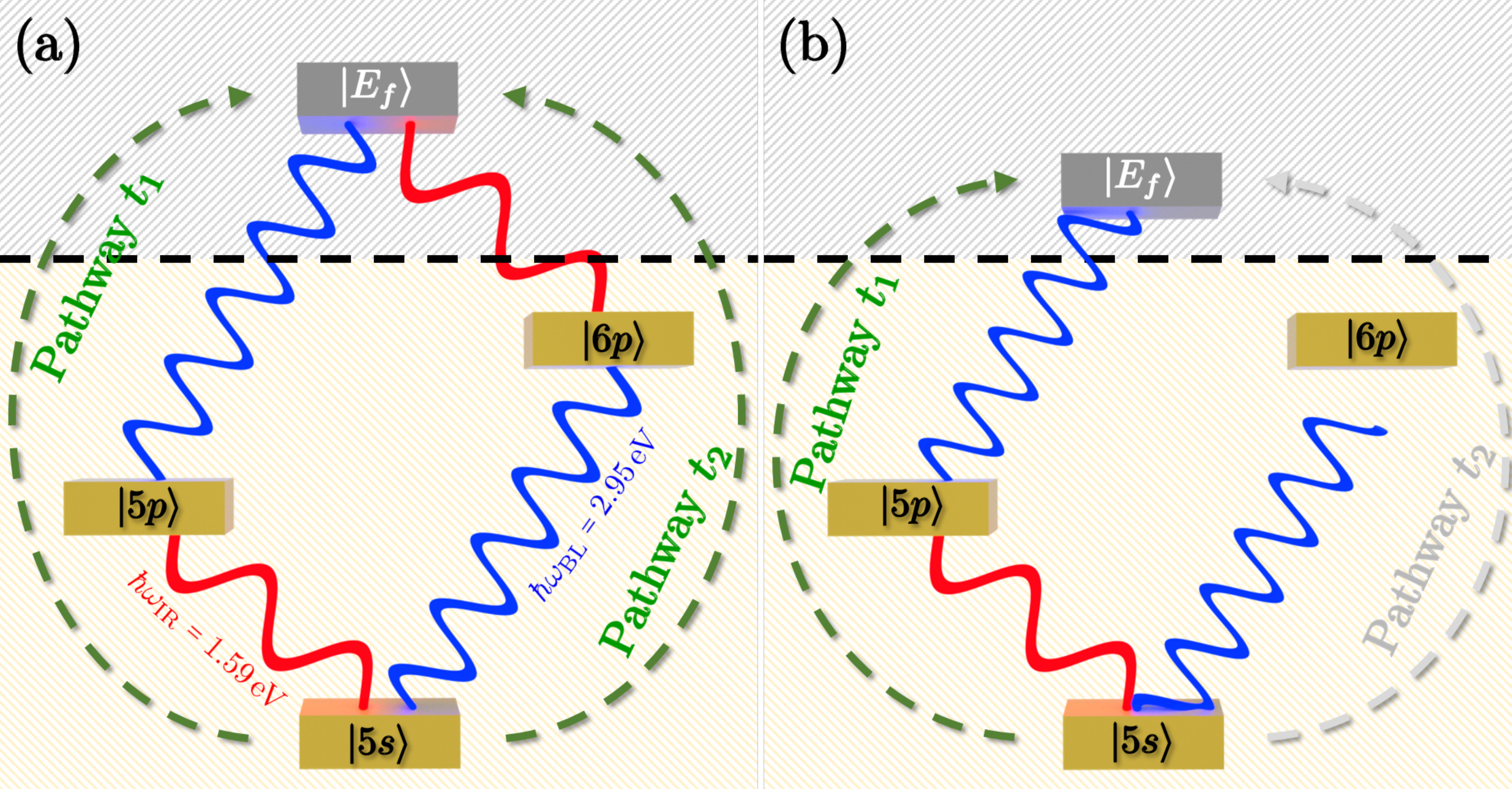}
\caption{(a) Schematic representation of the atomic "double slit
experiment" in Rubidium. The slits are represented by the bound and
initially unoccupied $5p$ and $6p$ intermediate states which are
resonantly excited by the infrared and blue laser pulses. The two
lasers subsequently ionize the intermediate levels, giving rise to two
ionization pathways $t_1$ and $t_2$. The photoelectron is transferred
into the continuum with a finite kinetic energy $E_f$ without knowing
which pathway was taken. (b) Closing one slit by detuning one of the
two lasers which then inhibits the occupation of the respective
intermediate state (in this example the $6p$ state). We hence
obtain a conventional two-color photoionization process via a single
ionization pathway. The energy level of the final state is now shifted down due
to the detuning decreasing effectively $\varepsilon_f=\hbar\omega_{\rm BL}+\hbar\omega_{\rm IR}+\varepsilon_{5s}$.}
\label{fig0}
\end{figure}

 In a typical double-slit experiment interference fringes are formed
on a screen placed behind the slits which are then traversed by
particles of suitable wavelength.  By blocking one of the slits, the
spatial interference pattern disappears.
 In a recent wave-vector space double slit experiment
\cite{pursehousedynamic} a photo-electron wave packet receding from
a single atom   is forced to pass through two energy levels within the atom, so that the levels play the role of the double-slit. As illustrated  in
Fig.\,\ref{fig0}, this is achieved by two low-intensity
continuous-wave lasers which  excite the $5p$ and $6p$ states of a
Rubidium atom.  The infrared light field can concurrently ionize the
$6p$ state and the blue laser the $5p$ state.   We image the interference in
wave-vector space by scanning the photoelectron angular distribution.
The interference pattern disappears if only one state is excited,
demonstrating  that the phase relationship between the interfering
waves is imprinted by the atom. The two-lasers are not phase locked.

%
The $5p$ and $6p$ states thus represent  "slits", which can be closed by
detuning the respective laser fields, leading to non-resonant
excitation and damped occupation  of the specific intermediate state.
This "damping" is however different from thermal damping,
as it does not swiftly destroy the coherence. The interference in
this case is affected because the two interfering amplitudes
have largely different strengths.

In the experiment the interference term and the associated
difference in the phases of the amplitudes  were  recorded by
essentially three measurements. First, both laser fields were set resonant
to the $5p$ and $6p$ dipolar excitations leading to an ionization
amplitude $t_1+t_2$ [cf.\,\ref{fig0}(a)]. In the second and third
measurements, one of the two laser fields was  detuned to
close one of the ionization pathways [cf.\,\ref{fig0}(b)] to extract
the individual amplitudes $t_1 (t_2)$. The ground state energy of the
$5s$ Rb state is -4.177\,eV while $E_{6p}=-1.589\,$eV and
$E_{5p}=-2.950\,$eV. Thus, no other states in the bound spectrum of
Rb are accessible by single-photon processes.  An interesting point is
that in the experiment the two \emph{cw} (continuous wave) lasers (cf. Fig.\ref{fig0}) were
not phase locked and were weak. The calculations show that a
random phase between these two lasers does not affect the
interference.  Thus the observed interference has to stem from the
atom, while the photoelectron wave is propagating out to the
detectors at infinity.

The question we are considering here is how the interference
behaves when the number of photons (the laser intensity) increases,
and hence one would expect an increase in the probability that a
blue and red photon are absorbed at the same time. The same process applies when
we consider  much shorter pulses. In this case one would expect the
phase relation between the laser pulses to be important, and it then becomes possible
to access the time scale on which the interference pattern  builds up and evolves.
 Unfortunately, due to experimental limitations we are currently not in a
position to investigate these ideas in the laboratory, and therefore the
current work is mostly theoretical.

In the final sections of this paper we discuss mechanisms for
controlling and manipulating the interference phenomena between both
ionization pathways. Atomic units are used throughout the paper.

\section{Propagation on a space-time grid}
\label{sec:2}

Within the single-particle picture the Rubidium atom is well
described by the angular-dependent model potential introduced by
Marinescu \emph{et al.} \cite{marinescu1994dispersion}:
\begin{equation}
V_\ell(r)=-\frac{Z_\ell(r)}{r} -
\frac{\alpha_c}{2r^4}\left[1-e^{-(r/r_c)^6}\right],
\end{equation}
where $\alpha_c$ is the static dipole polarizability of the
positive-ion core and the effective radial charge $Z_\ell(r)$ is
given by
\begin{equation}
Z_\ell(r)=1+(z-1)e^{-a_1r}-r(a_3+a_4r)e^{-a_2r},
\end{equation}
with the nuclear charge $z$ and the cut-off parameter $r_c$ as well
as the parameters $a_1-a_4$ fitted to experimental values. The
optimized parameters are tabulated in
Ref.\,\cite{marinescu1994dispersion}.

On a very fine space grid ($\Delta r=0.005$\,a.u.) the radial wave
functions of the atomic eigenstates
${\langle\pmb{r}|\phi_i\rangle=R_{n_i,\ell_i}(r)Y_{\ell_i,m_i}(\Omega_{\pmb{r}})}$
are found by (numerical) diagonalization of the matrix corresponding
to the time-independent Hamiltonian
$\hat{H}_0=-(1/2)\partial_r^2+{\ell(\ell+1)/(2r^2)} + V_\ell(r)$. In
the presence of solenoidal and moderately intense electromagnetic
fields the light-matter interaction Hamiltonian is given by $\hat{H}_{\rm
int}(t)=-\pmb{A}(\pmb{r},t)\cdot\hat{\pmb{p}}$ where
$\pmb{A}(\pmb{r},t)$ is the vector potential and $\hat{\pmb{p}}$ is
the momentum operator.

To account for all multi-photon and multipole effects, a numerical
propagation of the ground state wave function in the external vector
potential is necessary, e.g. by the matrix iteration scheme
\cite{nurhuda1999numerical}. Exploiting the spherical symmetry of the
atomic system, the time-dependent wave function is decomposed in
spherical harmonics, i.e. $\Psi(\pmb{r},t)=\sum_{\ell,m}^{\ell_{\rm
max}}b_{\ell,m}(r,t)Y_{\ell,m}(\Omega_{\pmb{r}})$. We therefore have to
propagate $(\ell_{\rm max} + 1)^2$ channel functions
$b_{\ell,m}(r,t)$ which are coupled through the corresponding
(dipole) matrix elements
$\langle\ell'm'|\pmb{A}\cdot\hat{\pmb{p}}|m\ell\rangle$. Initially,
the ground state channel is fully occupied by the 5s Rubidium
orbital, i.e. ${\Psi(\pmb{r},t\rightarrow-\infty)=
\langle\pmb{r}|\phi_{5s}\rangle}$ meaning
${b_{0,0}(r,t-\infty)=R_{n_i=5,\ell_i=0}(r)}$. Introducing a time
$T_{\rm obs}$ where the external electromagnetic field perturbation
is off, the wave function $\Psi(\pmb{r},t)$ is propagated to a
time $t>T_{\rm obs}$ where the photoelectron wave packet is fully
formed. The radial grid is extended to $10^4$\,a.u. to avoid
nonphysical reflections at the boundaries. Additionally, we
implemented absorbing boundary conditions by using an imaginary
potential. The resulting simulation shows that the
electron density at the final grid point $r_N$ is then smaller than the
numerical error at the considered propagation times. \\
To obtain the scattering properties of the liberated electron, we
project  $\Psi(\pmb{r},T_{\rm obs})$ onto a set of continuum wave
functions which are given by the partial wave decomposition:
\begin{equation}
\langle\pmb{r}|\varphi_{\pmb{k}}^{(-)}\rangle=\sum_{\ell,m}i^\ell
R_{k\ell}(r)e^{-i\delta_\ell(k)}
Y^*_{\ell,m}(\Omega_{\pmb{k}})Y_{\ell,m}(\Omega_{\pmb{r}}).
\label{eq:continuum}
\end{equation}
Here, the kinetic energy is defined by $E_k = k^2/2$,
$\delta_\ell(k)$ are the scattering phases and $R_{k\ell}(r)$ are
radial wave functions satisfying the stationary radial Schrödinger
equation for positive energies in the same pseudopotential
$V_\ell(r)$ which is used for obtaining the bound spectrum. The
scattering phases $\delta_\ell(k)={\rm
arg}[\Gamma(1+\ell-i/k)]+\eta_\ell(k)$ consist of the well-known
Coulomb phases (first term) and phase $\eta_\ell(k)$ characterizing
the atomic-specific short-ranged deviation from the Coulomb
potential. The radial wave functions are normalized to $\langle
R_{k\ell}|R_{k'\ell}\rangle= \delta(E_k-E_{k'})$. Finally, the
projection coefficients are given by
\begin{equation}
\begin{split}
a_{\ell,m}(k) =&\,e^{i(E_kT_{\rm obs} + \delta_\ell(k) - \ell\pi/2)}
\\
&\times\int_{r>r_a}{\rm d}r\,b_{\ell,m}(r,T_{\rm obs})R_{k\ell}(r).
\end{split}
\end{equation}
Here, we introduce the core radius $r_a$ and ensure that the
integration region is outside the residual ion. The photoionization probability (differential cross
section, abbreviated as DCS in the following) is defined as
\begin{equation}
\begin{split}
{\rm DCS} = \frac{{\rm d}\sigma}{{\rm d}\Omega_{\pmb{k}}}(E_k,\Omega_{\pmb{k}})
\propto &\sum_{\ell,\ell'}\sum_{m,m'} a^*_{\ell',m'}(k)a_{\ell,m}(k)
\\ &\times
Y^*_{\ell',m'}(\Omega_{\pmb{k}})Y_{\ell,m}(\Omega_{\pmb{k}}).
\end{split}
\end{equation}
while the total cross section is
$\sigma(E_k)\propto\sum_{\ell,m}\sigma_{\ell,m}(E_k)$ with
$\sigma_{\ell,m}(E_k)=|a_{\ell,m}(k)|^2$. Note that this treatment
gives explicit insight into the population of the individual angular
channels and their contributions to the photoelectron wave packet.

In the following two-color ionization of Rubidium, the electric
fields of both pulses are modeled according to
\begin{equation}
E_p(t) =
\epsilon_p\mathcal{E}_p\Omega(t+\Delta_p)\cos(\omega_p(t+\Delta_p)+\phi_p)
\end{equation}
with $p = {\rm IR,BL}$ standing for the infrared and blue laser
pulses, respectively. The polarization vectors are $\epsilon_p$, the
temporal envelope is given by $\Omega(t)=\cos^2(\pi t/T_d^p)$ for
$t\in[-T_d^p/2,T_d^p/2]$ with the pulse duration $T_d^p=2\pi
n_p/\omega_p$
determined by the number of optical cycles $n_p$. Further we
introduce a temporal difference $\Delta_p$ and a phase difference
$\phi_p$ to account for both laser fields originating from
different sources, and which are hence not phase-locked. Without loss of
generality we set $\phi_{\rm IR}=0$ and $\Delta_{\rm IR}=0$. Both
laser fields are assumed to be linearly polarized in the $z$-direction
so that the
azimuthal angular quantum number $m$ is conserved. The number
of angular channels then reduces to $\ell_{\rm max}+1$ and in the
following treatment we omit the subscript $m$ for brevity.
\begin{figure}[t]
\includegraphics[width=1.0\columnwidth]{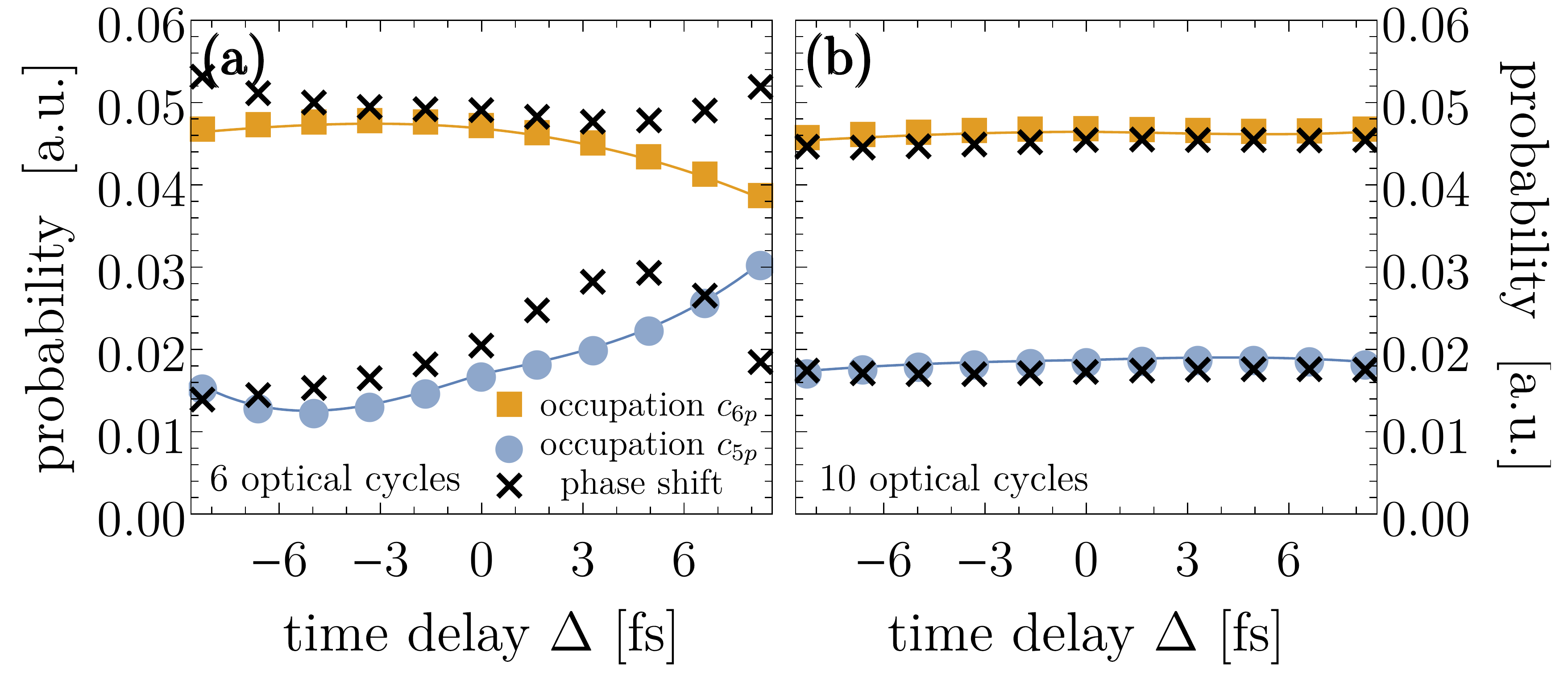}
\caption{Occupation numbers of the $5p$ and $6p$ intermediate states
after laser excitation, which depend on the time delay $\Delta$ and
the phase difference  between both pulses. (a) shows the results of
the numerical propagation for six optical cycles. (b) corresponds to
ten optical cycles. Dots indicate the results for $\phi=0$ radians and
crosses belong to $\phi=2/3\pi$.}
\label{fig1}
\end{figure}
To balance the differences in the oscillator strengths between the
$5s\rightarrow5p$ and $5s\rightarrow6p$ channels we used the field strengths
$\mathcal{E}_{\rm BL}=0.05$ a.u. and $\mathcal{E}_{\rm IR}=0.007$
a.u. in the following simulations.

The aim here is to determine the influence of the phase
differences on the resulting photoionization and occupation
probabilities. In Fig.\,\ref{fig1} we present the occupation numbers
of the intermediate $5p$ and $6p$ states for two different laser
configurations at a time $T_{\rm obs}$ after laser excitation, meaning both
light pulses are completely extinguished and the photoelectron wave packet
propagates freely in the Coulomb field of the residual Rubidium ion.
Panel (a) corresponds to the case where both laser pulses have a
length of six optical cycles. We see that both occupation numbers
show a strong dependence on the temporal difference $\Delta$ as well
as on a random phase difference $\phi$. The dots belong to $\phi=0$
while crosses indicate the results for a non-zero phase difference.
Here, we show the occupation numbers for $\phi=2/3\pi$, which show a
large difference to the case of $\phi=0$. In addition, we repeated
the simulation for other numbers of the random phase difference and
obtained similarly pronounced discrepancies. The situation changes
completely when increasing the number of optical cycles as shown in
panel (b). For ten optical cycles the influence of both
the temporal and phase differences on the resulting occupation
numbers is drastically reduced, pointing to the transition into the
\emph{cw} limit.

\begin{figure}[t]
\includegraphics[width=1.0\columnwidth]{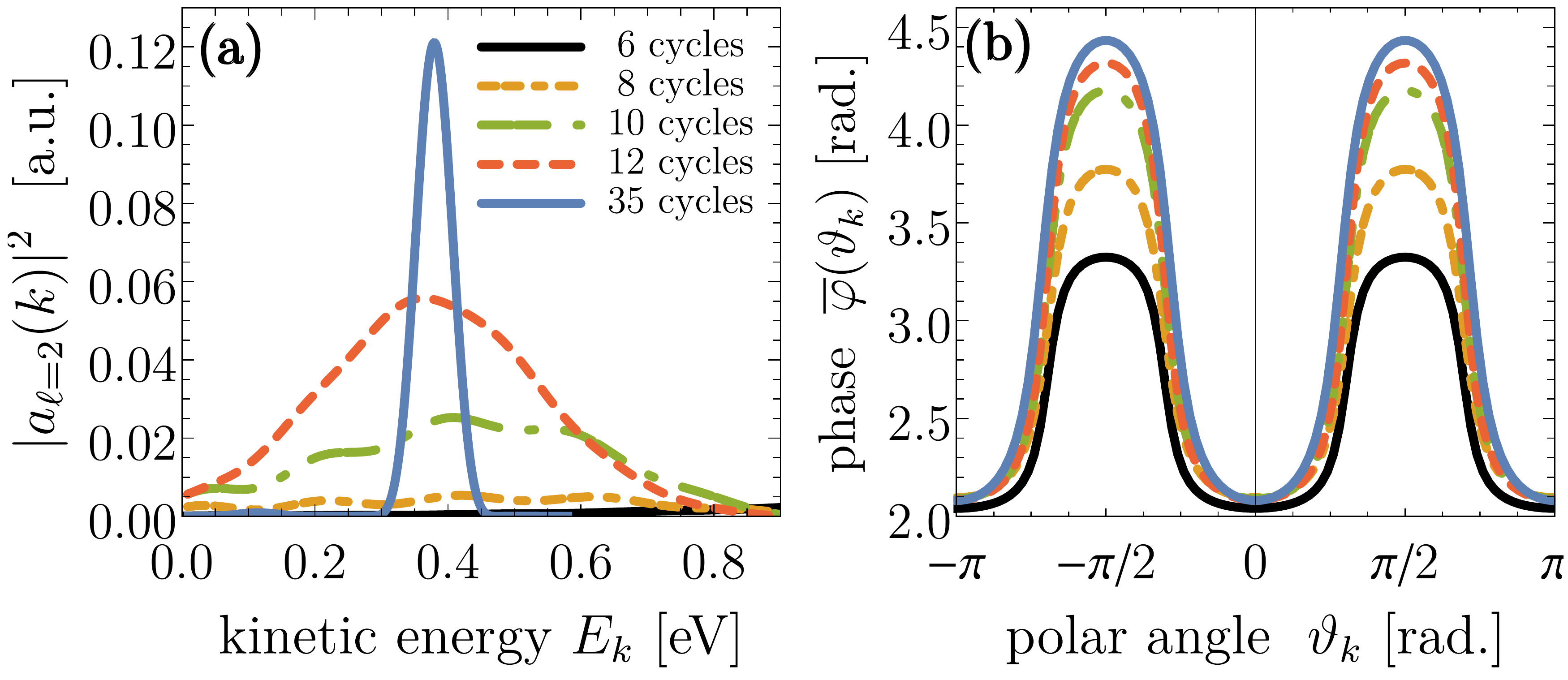}
\caption{Properties of the ejected photoelectron. (a) Ionization
probability of the angular channel $\ell=2$ for different pulse
lengths of the incident IR and blue laser fields. (b) Angular
variation of the averaged quantum phase
$\overline{\varphi}(\vartheta_k)$ in dependence on the number of
optical cycles.}
\label{fig2}
\end{figure}

Already from the \emph{bounded} properties we suspect the rather fast
convergence of the photoionization process into a description
within the frequency domain, which is characterized by infinitely long
laser pulses characterised by delta distribution-like bandwidths. We can
underline this observation by looking at the characteristics of the
ejected photoelectron wave packet. In Fig.\,\ref{fig2}(a) we present
the ionization probability of the angular channel $\ell=2$
characterized by the partial cross section
$\sigma_2=\left|a_{\ell=2}(k)\right|^2$. As expected, the probability
curve sharpens under an increase in the pulse lengths of both laser
fields. While for $n_p=6$ a manifold of energy states in the
continuum is excited, from $n_p=10$ upwards we see clearly the
unfolding of a Gaussian-like peak around the final energy
$E_f=E_{5s}+\hbar\omega_{\rm IR} + \hbar\omega_{\rm BL}$. In the case
of 35 optical cycles the FHWM of the probability peak is around
0.07\,eV.

It is also interesting to study the evolution of the quantum phase associated with
the photoelectron wave packet, which can be expressed as
$\varphi(k,\vartheta_k)={\rm
arg}\left[\sum_{\ell}a_\ell(k)Y_{\ell,0}(\Omega_{\pmb{k}})\right]$.
For 6 optical cycles, the angular channels with $\ell=0$ and
$\ell=2$ already represent the dominant contributions to the photoelectron
at the final energy around 0.4 eV, as expected for a two-color
photoionization process of an initial $s$-state. The quantum phase is
a result of interference between both partial waves and for the long
pulse limit it may be mathematically expressed by
\cite{watzel2014angular}
\begin{equation}
\varphi(k,\vartheta_k)=\arctan\left[\frac{\sum_{\ell=0,2}S_\ell(k,\vartheta_k)\sin(\phi_\ell(k))}{\sum_{\ell=0,2}
S_\ell(k,\vartheta_k)\cos(\phi_\ell(k))}\right]
\label{eq:phase}
\end{equation}
with
$S_\ell(k,\vartheta_k)=\left|a_\ell(k)\right|Y_{\ell,0}(\Omega_{\pmb{k}})$
and $\phi_\ell(k)={\rm arg}[a_\ell(k)]\simeq\delta_\ell(k)-\ell\pi/2$.
The quantum phase hence depends crucially on the scattering
phases $\delta_\ell(k)$ and on the ratio between the transitions
strengths into $\ell=2$ and $\ell=0$ angular channels respectively.
We note that in principle $\varphi(k,\vartheta_k)$ is an experimentally accessible
quantity, since it can be recovered by integrating the Wigner time
delay in photoionization defined as $\tau_{\rm
W}(E_k,\vartheta_k)=({\rm d}/{\rm d}E_k)\varphi(k,\vartheta_k)$.
This can be extracted from delay measurements that are possible due
to recent experimental advances within the attosecond timeframe
\cite{schultze2010delay, isinger2017photoionization}. The measured atomic time delay $\tau_{\rm a}$ consists of
an "intrinsic" contribution (Wigner time delay $\tau_{\rm
W}$) upon the absorption of an XUV photon which can be interpreted
as the group delay of the outgoing photoelectron wave packet due to the collision process. As mentioned above, it contains information about the internal quantum phase. The second term $\tau_{\rm cc}$ arises from continuum-continuum transitions due to the interaction of the laser probe field
with the Coulomb potential and depends crucially on the experimental parameters. Hence, the difference $\tau_{\rm W}=\tau_{\rm a} - \tau_{\rm cc}$ provides access to the phase information $\varphi(k,\vartheta_k)$.

In Fig.\ref{fig2}(b) it is shown how the
phase develops by increasing the pulse lengths (number of optical
cycles). Here, we show the phase averaged over the ionization probability (total cross section
$\sigma(E_k)$): $\overline{\varphi}(\vartheta_k)=\int{\rm
d}E_k\,\sigma(E_k)\varphi(k,\vartheta_k)/\int{\rm
d}E_k\,\sigma(E_k)$. Interestingly for very short pulses
($n_p=6$) where the cross section is far from being centered
around a final energy $E_f=E_{5s}+\hbar\omega_{\rm IR} +
\hbar\omega_{\rm BL}$, the shape of the phase matches that extracted
from simulations with longer pulses. As anticipated from earlier
results, from ten cycles upwards the results converge quickly into the
\emph{cw} limit. As an example, this is seen since the discrepancy between the quantum
phase for 12 and 35 optical cycles is smaller than 5\%.

\section{From short pulses to the \emph{cw}-limit}
\label{sec:3}

\begin{figure*}[t!]
\includegraphics[width=0.95\textwidth]{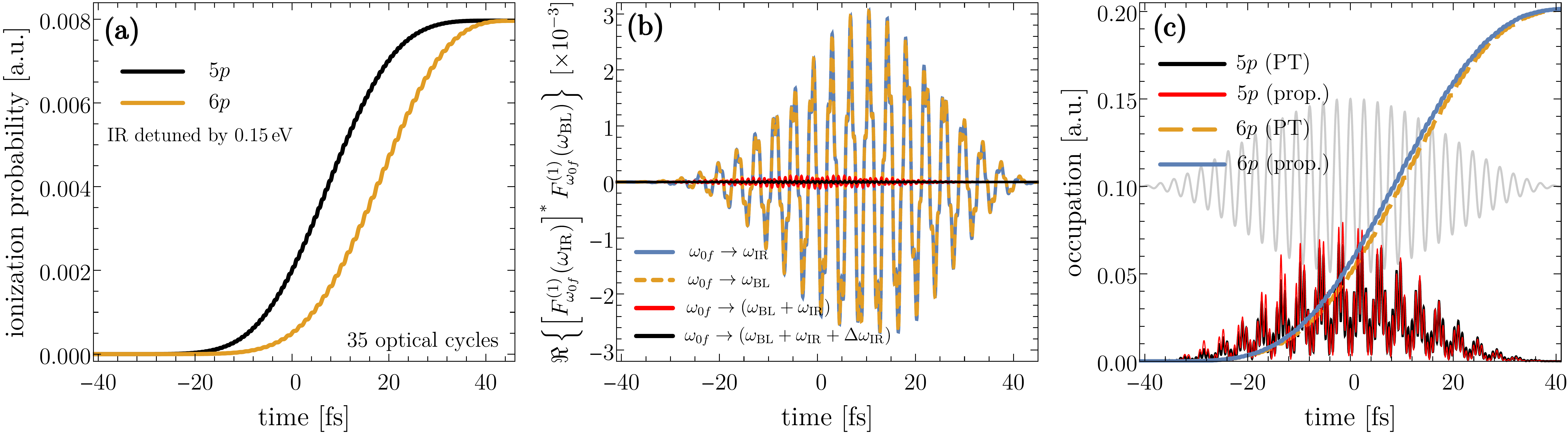}
\caption{(a) Comparison between photoionization pathways via $5p$ and
$6p$ intermediate states in the case of detuning $\delta\omega_{\rm
IR}=0.15$\,eV of the infrared field. The curves are extracted from
Eq.\,\eqref{eq:d2_2}. (b) Time dependence of the first-order product
$\Re\{[F_{\omega_0}^{(1)}(t,-\omega_{\rm
IR},0)]^*F_{\omega_0}^{(1)}(t,-\omega_{\rm BL},\Delta_{\rm BL})\}$
for different final energies. (c) Time dependent occupation numbers
$c_{5p}(t)$ and $c_{6p}(t)$ extracted from perturbation theory and
a full numerical treatment. The grey curve indicates the temporal
variation of the IR laser field.}
\label{fig3}
\end{figure*}

By considering the two-color ionization process using perturbation
theory, we express the time-dependent wave function as
$\Psi(\pmb{r},t)=\sum_{\nu\ell} d_{\nu\ell}(t)e^{-{\rm
i}E_{\nu\ell}t}\phi_{\nu\ell}(\pmb{r})$. Note that the quantum number
$\nu$ includes both the bound and the continuum states. The first
order amplitude is given by:
\begin{equation}
\begin{split}
d^{(1)}_{0\rightarrow
f}(t)=&-\frac{1}{i}\sum_{\lambda=\pm1}\left[\langle f|D_{\rm
IR}|n\rangle \mathcal{E}_{\rm
IR}F_{\omega_{0f}}^{(1)}(t,\lambda\omega_{\rm IR},\Delta=0) \right. \\
&\left.+ e^{i\lambda\Delta\omega_{\rm BL}} \langle f|D_{\rm
BL}|n\rangle \mathcal{E}_{\rm
BL}F_{\omega_{0f}}^{(1)}(t,\lambda\omega_{\rm
BL},\lambda\Delta)\right],
\end{split}
\label{eq:d1}
\end{equation}
where $D_i=\hat{\epsilon}_i\cdot\hat{d}$ ($i={\rm IR,BL}$) is the
dipole operator, $\omega_{0f}=E_f-E_0$ and
\begin{equation}
F_{\omega_{0f}}^{(1)}(t,\omega,\Delta)= \int_{-\infty}^{t}
\Omega(t'+\Delta)e^{{\rm i}(\omega_{0f}+\omega)t'} {\rm d}t'.
\end{equation}
Without loss of generality we assume both laser fields are
described by the same temporal function $\Omega(t)=\cos^2[\pi t/T_p]$
so that both pulses have the same pulse length
$T_p=2n_p\pi/\omega_{\rm IR}$. In the following discussion the number of
optical cycles $n_p$ hence refers to the infrared laser field. For the
squared-cosine shaped envelope $\Omega(t)$ of the pulses, the
function $F_{\omega_{0}}^{(1)}(t,\omega,\Delta)$ can be obtained
analytically and converges against
$F_{\omega_{0}}^{(1)}(t,\omega,\Delta)\rightarrow\delta(\omega_{0}-\omega)$
for $T_p\rightarrow\infty$. The second order amplitude yields the
following expression:
\begin{equation}
\begin{split}
d^{(2)}_{0\rightarrow
f}(t)=&-\sum_{i,j}\sum_{\lambda,\lambda'=\pm1}\sum_n
e^{{\rm i}(\lambda\Delta_j\omega_j+\lambda'\Delta_i\omega_i)}\\
&\times\mathcal{E}_j\mathcal{E}_i \frac{\langle f|D_j|n\rangle
\langle n|D_i|0\rangle}{\omega_{0n}-\omega_i} \\
&\times
F^{(2)}_{\omega_{nf},\omega_{0n}}(t,\lambda\omega_j,\lambda'\omega_i,\lambda\Delta_j,\lambda'\Delta_i),
\end{split}
\label{eq:d2_2}
\end{equation}
where again $i,j={\rm IR,BL}$. The second-order temporal function is
defined as
\begin{equation}
\begin{split}
F&^{(2)}_{\omega_{nf},\omega_{0n}}(t,\omega_j,\omega_i,\Delta_j,\Delta_i)=(\omega_{0n}-\omega_i)\\
&\times\int_{-\infty}^{t}{\rm d}t'\,\Omega(t'+\Delta_j)e^{{\rm
i}(\omega_{nf}+\omega_j)t'}F^{(1)}_{\omega_{0n}}(t',\omega_i,\Delta_i).
\end{split}
\end{equation}
A closed expression for the second-order temporal function $F^{(2)}$
cannot be obtained analytically. However, a solution for $t>T_p/2$
(time of switch off) can be found and investigated for
$T_p\rightarrow\infty$ (the continuous wave limit). It follows that in
the many optical cycle limit we find energy conservation, i.e.
\begin{equation}
\lim_{n_p\rightarrow\infty}
F^{(2)}_{\omega_{nf},\omega_{0n}}(t,-\omega_j,-\omega_i,\Delta_j,\Delta_i)
= \frac{3\pi}{4}\delta(\omega_{0f} - \omega_{\rm j} - \omega_{\rm i})
\label{eq:F2_asymptotic}
\end{equation}
Further, in this limit the individual temporal shifts $\Delta_j$ and
$\Delta_i$ have no influence on the (second-order) transition
amplitude. Thus, in the long (\emph{cw}) pulse limit and for energies around
$E_f=E_0+\hbar\omega_{0f}$, this behavior allows us to rewrite the
resulting second-order transition amplitude:
\begin{equation}
\begin{split}
d^{(2)}_{0\rightarrow
f}\underset{n_p\rightarrow\infty}{\rightarrow}&\frac{3\pi}{4}\mathcal{E}_{\rm
IR}\mathcal{E}_{\rm BL}e^{{\rm i}\Delta\omega_{\rm
BL}}\sum_n\left[\frac{\langle f|D_{\rm IR}|n\rangle\langle n|D_{\rm
BL}|0\rangle}{\omega_{n0}-\omega_{\rm BL}}\right. \\
&\left.+ \frac{\langle f|D_{\rm BL}|n\rangle\langle n|D_{\rm
IR}|0\rangle}{\omega_{n0}-\omega_{\rm IR}} \right],
\end{split}
\label{eq:2PhotonMatrix}
\end{equation}
which is similar to the traditional form of the two-photon matrix
element \cite{toma2002calculation}. From here we see directly that a
random phase $\phi=\Delta\omega_{\rm BL}$ does not play any role
(especially when analyzing the cross sections
$\sim|d^{(2)}_{0\rightarrow f}|^2$.

Let us now come back to the two-color ionization process in the Rubidium
atom. It is crucial for the first and second-order amplitudes to
precisely evaluate the dipole matrix elements \cite{amusia2013atomic}
\begin{equation}
\langle
f|D_i|n\rangle=(-1)^{m_f+\ell_>}\begin{pmatrix}\ell_f&1&\ell_n \\
-m_f&0&m_n \end{pmatrix}\sqrt{\ell_>}\langle f||\hat{d}||n\rangle.
\end{equation}
where $\ell_>={\rm max}(\ell_f,\ell_n)$ and the reduced radial matrix
element $d_{\ell_f\ell_n}=\langle
f||\hat{d}||n\rangle=\int_0^\infty{\rm d}r\,R_f(r)D_iR_n(r)$. 
One way to account, at least partially, for the the interaction between the valence  and the core electrons \cite{hameed1968core}
 is to modify the operator  $\hat{Q}_L$ as 
\begin{equation}
\hat{Q}_L\rightarrow\hat{Q}_L\left[1-\frac{a_c^{(L)}}{r^{2L+1}}\left(1-e^{-(r/r_c')^{2L+1}}\right)\right],
\label{eq:core_Q}
\end{equation}
where $a_c^{(L)}$ is the  $2^L$ tensor core polarizability and $r_c'$ is an empirical cut-off radius (for Rb $r_c'=4.339773$ a.u.
\cite{marinescu1994dispersion}). This physical picture behind the corrections is roughly that  the valence electron with a dipole moment $\pmb{d}$ induces  (by virtue of its field)  a (core) dipole moment $-\alpha_c\pmb{d}/r^3$ (and higher multipoles).  Then, the complete dipole moment becomes $\pmb{d}(1-\alpha_c/r^3)$. Note that in our case the Dipole operator $\hat{d}=\hat{Q}_1$. Using the modified dipole operator delivers very accurate matrix elements near the ionization threshold in comparison with experiments and more sophisticated theoretical models
\cite{petrov2000near}.

We learn from Eq.\,\eqref{eq:2PhotonMatrix} that changing the
electric field amplitudes $\mathcal{E}_{\rm IR}$ and
$\mathcal{E}_{\rm BL}$ will not balance any differences between the
matrix element products $\langle f|D_{\rm IR}|n\rangle\langle
n|D_{\rm BL}|0\rangle$ and $\langle f|D_{\rm BL}|n\rangle\langle
n|D_{\rm IR}|0\rangle$. Thus, to reach equipollent ionization
pathways $E_{5s}\rightarrow E_{5p}\rightarrow E_f$ and
$E_{5s}\rightarrow E_{6p}\rightarrow E_f$ we have to detune the laser
frequency corresponding to the stronger bound-bound transition. Given
the reduced matrix elements $\langle 5p||\hat{d}||5s\rangle=-5.158$
and $\langle 6p||\hat{d}||5s\rangle=0.468$, we have to detune the
infrared field as shown in Fig\,\ref{fig3}(a). The curves are
obtained from a numerical integration of Eq.\,\eqref{eq:d2_2}. For
$n=35$ optical cycles, a detuning of
$\delta\omega_{\rm IR}=+0.15$\,eV is required to allow the second-order
transition amplitudes within each pathway to have the same magnitude.
Note that the value of $\Delta\omega_{\rm IR}$ decreases by
increasing the pulse lengths. In this vein we performed the
simulation for 75 optical cycles and found that a detuning of only
0.07 eV is needed to reach equipollent ionization pathways (not shown
for brevity).

Let us consider the time-dependent first order probability
$P_{0\rightarrow f}^{(1)}(t)=\left|d^{(1)}_{0f}(t)\right|^2$ which
reads explicitly
\begin{equation}
\begin{split}
P_{0\rightarrow f}^{(0)}(t)=&\left|\langle f|D_{\rm IR}|n\rangle
\mathcal{E}_{\rm IR} F_{\omega_{0f}}^{(1)}(t,-\omega_{\rm
IR},0)\right|^2 \\
&+ \left|\langle f|D_{\rm BL}|n\rangle \mathcal{E}_{\rm
BL}F^{(1)}_{\omega_{0f}}(t,-\omega_{\rm BL},\Delta_{\rm BL})\right|^2
\\
&+2\mathcal{E}_{\rm IR}\mathcal{E}_{\rm BL} \langle f|D_{\rm
IR}|n\rangle\langle f|D_{\rm BL}|n\rangle \\
&\times\Re\left\{F_{\omega_{0f}}^{(1)}(t,-\omega_{\rm
IR})\left[F_{\omega_{0f}}^{(1)}(t,-\omega_{\rm BL},\Delta_{\rm
BL})\right]^*\right\}.
\end{split}
\end{equation}
Here, we have to emphasize that all terms containing $+\omega_i$ are
negligibly small which we refer to as the rotating wave approximation.
The last term in the third line might look like a two-photon process
but a closer inspection of the function
$F_{\omega_0}^{(1)}(t,-\omega,\Delta)$ reveals that it sharply peaks
around $\omega_0-\omega=0$ even for times close to $T_p/2$. Since
$\omega_{f0}$ is fixed, the product between both $F^{(1)}$ functions
is zero, so that
\begin{equation}
\begin{split}
\lim_{T_p\rightarrow\infty} F_{\omega_{f0}}^{(1)}&(T_p/2,-\omega_{\rm
IR},0)\left[F_{\omega_{f0}}^{(1)}(T_p/2,-\omega_{\rm BL},\Delta_{\rm
BL})\right]^* \\
&= \delta(\omega_{f0}-\omega_{\rm BL})\delta(\omega_{f0}-\omega_{\rm
BL}).
\end{split}
\end{equation}
As confirmation, in Fig.\,\ref{fig3}(b) we show the time-dependent
product of the functions $F_{\omega_0}^{(1)}(t,-\omega_{\rm IR},0)$
and $F_{\omega_0}^{(1)}(t,-\omega_{\rm BL},\Delta_{\rm BL})$ for
different $\omega_{f0}$. As stated above, all situations have in
common that the product is zero at the time when the pulse is
switched off, representing energy conservation. Finally, the
first-order transitions for $t=T_p/2$ are given by
\begin{equation}
\begin{split}
d^{(1)}_{0\rightarrow f}(t>T_p/2)=&-\frac{1}{2i}\left[\mathcal{E}_{\rm
IR}\langle f|D_{\rm IR}|0\rangle\delta(w_{f0}-\omega_{\rm IR})\right.
\\
&\left.+ e^{{\rm i}\Delta\omega_{\rm BL}}\mathcal{E}_{\rm BL}\langle
f|D_{\rm BL}|0\rangle\delta(w_{f0}-\omega_{\rm BL})\right].
\end{split}
\end{equation}
The energies of the blue and red photons are not sufficiently high to
reach the continuum and only the two-photon matrix element developed
in Eq.\,\eqref{eq:d2_2} gives insight into the properties of the
photoelectron.

As a consequence, the amplitude $d^{(1)}$ describes the
photoexcitation process of the intermediate $f=5p$ and $f=6p$ states.
However, due to the sharp laser pulses the same final state $f$
cannot be excited by both photons. Hence, in the resulting
photoexcitation probabilities $c_f=|d^{(1)}_{0\rightarrow
f}(t>T_p/2)|^2$ the random phase $\phi=\Delta\omega_{\rm BL}$ does
not play a role, which confirms the full-numerical results shown in
Fig.\,\ref{fig1}(b).
\begin{figure*}[t!]
\includegraphics[width=0.95\textwidth]{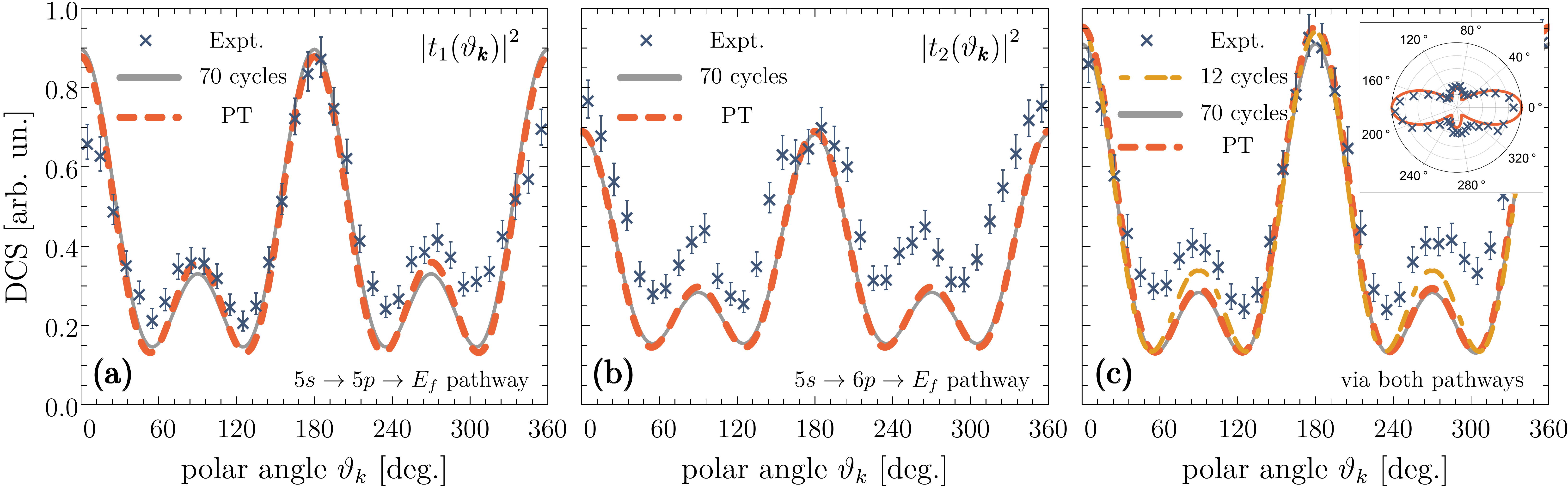}
\caption{(a) Differential cross section (DCS) when the blue laser field
is off-resonance, thereby effectively closing the second ionization pathway.
Comparison between two theoretical models (PT:red and
propagation:gray) and experimental results. For the propagation, the
blue field was detuned by $\delta\omega_{\rm BL}=0.13$\,eV. Further,
the infrared field was slightly detuned by $\delta\omega_{\rm
IR}=0.04$\,eV to manipulate the ratio between both two-photon
amplitudes (see discussions in Sec.\,\ref{sec:3}). (b) Same as in (a)
with the infrared field completely off-resonance, effectively closing
the first ionization pathway. The detuning amount is $\delta\omega_{\rm
IR}=0.14$\,eV. (c) Both laser fields resonant with associated
pathways. To achieve pathway amplitudes that are comparable with experiment
the infrared field was slightly detuned by $\delta\omega_{\rm
IR}=0.04$\,eV. The inset shows a polar plot of the DCS. All
theoretical curves are scaled by the same factor and are shifted by a
constant offset of 0.1 to aid  comparison with the experimental data.
The data points are reproduced from Ref.\cite{pursehousedynamic}.}
\label{fig4}
\end{figure*}
In panel \ref{fig3}(c) we present the occupation numbers
$c_{5p}(t)=|\langle\phi_{5p}|\Psi(t)\rangle|^2$ and
$c_{6p}(t)=|\langle\phi_{6p}|\Psi(t)\rangle|^2$ extracted from the
perturbative treatment (PT) in Eq.\eqref{eq:d1} and the numerical
simulation scheme developed in Sec.\,\ref{sec:2}. We find a remarkable
agreement between the PT results and the full numerical treatment,
demonstrating the transition into the \emph{cw} limit and the
validity of the perturbative treatment of the two-color ionization
problem. Due to the detuning of the IR field the $c_{5s}(t)$
term decreases at the end of the pulse while the $c_{6p}(t)$ term belongs to
resonant excitation ($5s\rightarrow6p$) in the blue laser field. This
is the reason for the nearly monotonous increase that is seen. We note that the corresponding
Rabi frequencies $\Omega_{5s-5p}$ and $\Omega_{5s-6p}$ are very small which means the occupation numbers in
Fig.\,\ref{fig3}(c) represent only the first segment of the first Rabi Cycle which can be identified by
the characteristic quadratic dependence on the time.

\section{Interference between both ionization pathways}
\label{sec:4}

As revealed by the two-photon matrix element in
Eq.\,\eqref{eq:2PhotonMatrix}, both laser fields act simultaneously
to produce the same final photoelectron state, and so we have to deal
with two transition amplitudes $t_1$ and $t_2$ which are presented by
the two terms. As already demonstrated in Sec.\,\ref{sec:2} the final
state $|f\rangle$ is mainly described by a superposition of two
angular channels $\ell=0$ and $\ell=2$. Thus, we may write
\begin{equation}
\begin{split}
d^{(2)}_{0\rightarrow
f}(\vartheta_{\pmb{k}})&= t_1(\vartheta_{\pmb{k}})+t_2(\vartheta_{\pmb{k}}) \\
&=\underbrace{S_{\ell=0}^{(t_1)}(\vartheta_{\pmb{k}})e^{i\phi_{\ell=0}(k_f)}
+
S_{\ell=2}^{(t_1)}(\vartheta_{\pmb{k}})e^{i\phi_{\ell=2}(k_f)}}_{t_1=|t_1|e^{i\varphi_1}}
\\
&\quad+\underbrace{S_{\ell=0}^{(t_2)}(\vartheta_{\pmb{k}})e^{i\phi_{\ell=0}(k_f)}
+
S_{\ell=2}^{(t_2)}(\vartheta_{\pmb{k}})e^{i\phi_{\ell=2}(k_f)}}_{t_2=|t_2|e^{i\varphi_2}}.
\end{split}
\end{equation}
The final state is in the continuum and is defined by
Eq.\,\eqref{eq:continuum}. Consequently, we can define
\begin{equation}
\begin{split}
S^{(t_i)}_{\ell_f}(\vartheta_{\pmb{k}})=&(-1)^{2+\ell_f/2}\sqrt{1+\ell_f/2}\begin{pmatrix}\ell_f&1&1
\\ 0&0&0 \end{pmatrix}\begin{pmatrix}1&1&0\\0&0&0 \end{pmatrix}\\
&\times\sum_n \frac{\langle E_f\ell_f||\hat{d}||n\rangle\langle
n||\hat{d}||5s\rangle}{\omega_{0n}-\omega_1} \\
&\times \mathcal{E}_{\rm IR}\mathcal{E}_{\rm BL}
F^{(2)}_{\omega_{nf},\omega_{0n}}(t\rightarrow\infty,-\omega_2,-\omega_1,0,0)\\
&\times Y_{\ell_f,0}(\vartheta_{\pmb{k}}).
\end{split}
\end{equation}
Here without loss of generality we set the time delays $\Delta_{\rm IR/BL}$ to
zero since we are in the \emph{cw} limit. For pathway $t_1$, $\omega_1=\omega_{\rm IR}$ and $\omega_2=\omega_{\rm
BL}$, while for pathway $t_2$ the opposite is required. Further,
$\phi_\ell(k)=\delta_\ell(k)-\ell\pi/2$ while the corresponding pathway phases $\varphi_{\rm i}$ are already defined
in Eq.\,\eqref{eq:phase}. Note, that $|t_1(\vartheta)|^2$ and
$|t_2(\vartheta)|^2$ define the DCS for the individual pathways while
the interference term between both pathways is given by
\begin{equation}
\begin{split}
{\rm DCS}_{\rm interf.} &=
t_1(\vartheta_{\pmb{k}})t_2^*(\vartheta_{\pmb{k}}) +
t_2(\vartheta_{\pmb{k}})t_1^*(\vartheta_{\pmb{k}}) \\
&=\left|t_1(\vartheta_{\pmb{k}})+t_2(\vartheta_{\pmb{k}})\right|^2 -
(\left|t_1(\vartheta_{\pmb{k}})\right|^2 +
\left|t_2(\vartheta_{\pmb{k}})\right|^2)
\end{split}
\end{equation}
The \emph{phase difference} related to ${\rm DCS}_{\rm interf.}$ is given by
\begin{equation}
\begin{split}
\Delta\varphi_{12}(\vartheta_{\pmb{k}})=\cos^{-1}\left[\frac{t_1(\vartheta_{\pmb{k}})t_2^*(\vartheta_{\pmb{k}})
+
t_2(\vartheta_{\pmb{k}})t_1^*(\vartheta_{\pmb{k}})}{2\left|t_1(\vartheta_{\pmb{k}})\right|
\left|t_2(\vartheta_{\pmb{k}})\right|}\right].
\end{split}
\end{equation}
The interference term is the result of differences in the ratios
$S^{(t_1)}_{\ell=2}/S^{(t_1)}_{\ell=0}$ and
$S^{(t_2)}_{\ell=2}/S^{(t_2)}_{\ell=0}$ between the $s$- and $d$-
partial waves associated with the individual pathways $t_1$ and
$t_2$. These ratios depend crucially on the reduced bound-continuum
dipole matrix elements $\langle E_f,\ell||\hat{d}||5p\rangle$ and
$\langle E_f,\ell||\hat{d}||6p\rangle$ as well as the bound-bound
dipole matrix elements $\langle 5p||\hat{d}||5s\rangle$ and $\langle
6p||\hat{d}||5s\rangle$.

The perturbative treatment of the two-pathway ionization process shares  some parallels with
the theoretical description of the recently developed attosecond measurement techniques \cite{dahlstrom2013theory}. For instance, the occurrence and spectral characteristics of the $2q$th sideband in the RABBITT scheme \cite{muller2002reconstruction} stem from the interference between two ionization pathways: the absorption of harmonic $H_{2q-1}$ or $H_{2q+1}$ plus absorption or emission of a laser photon with $\hbar\omega$. Hence, similarly to the effects studied in this work, the measured intensity of the side band  depends on the phase difference between the quantum paths. In contrast to typical attosecond experiments, we create a bound wave packet upon absorption of the first photon. This case was studied in photoionization of Potassium where the spectral properties of the initially created bound wave packet was used to eliminate the influence of the dipole phase in the angle-integrated photoelectron spectrum making it possible to fully characterize the attosecond pulses \cite{pabst2016eliminating}. Similar to the investigated experiment by Pursehouse and Murray, the underlying physical principle is the quantum interference of pathways corresponding to ionization from different energy levels. Moreover, a realistic many-body treatment revealed that correlation effects have only a minor influence in Alkali atoms which supports our theoretical treatment in this work.

In Fig.\,\ref{fig4} we present the individual $t_1$ DCS (a), $t_2$
DCS (b) and the DCS corresponding to the coherent summation $t_1+t_2$
(c). In all panels we compare the ionization probabilities extracted
from the full numerical and perturbative treatment with experimental
results from measurements performed by Pursehouse and Murray
\cite{pursehousedynamic}. In the experiment the individual pathway
cross sections are obtained by the appropriate  detuning of the
respective laser fields: To extract $|t_1|^2$
($5s\rightarrow5p\rightarrow E_f$) the blue laser beam is detuned to block
occupation of the $6p$ state. To obtain the $|t_2|^2$ amplitude
($5s\rightarrow6p\rightarrow E_f$), the infrared laser field is tuned
to be off-resonant to the $5p$ transition. To obtain the total
amplitude $|t_1+t_2|^2$ both laser pulses are on resonance to the
respective $5s\rightarrow np$ transitions. As explained in
Sec.\,\ref{sec:3} the infrared laser field is always slightly detuned by
a fixed $\delta\omega_{\rm IR}$ so that both two-photon
amplitudes $t_1$ and $t_2$ are of the same magnitude.

In panel \ref{fig4}(a) we show the DCS of the individual pathway 1.
For the full numerical treatment with a number of 70 optical cycles
we used
a blue detuning of $\delta\omega_{\rm BL}=0.13$\,eV while the red
field detuning amounts to $\delta\omega_{\rm IR}=0.04$\,eV. In the
perturbative treatment ($n_p\rightarrow\infty$) we need a much
smaller detuning (in the range of meV). Both theoretical models agree
extraordinary well with the experiment in general, while minor discrepancies can be found
around the maxima. A possible explanation is the shift of the final
energy by nearly 0.17\,eV in the full-numerical propagation (finite
number of optical cycles) due to the detuning of both fields which
changes slightly the ratios $S^{(t_1)}_{\ell=2}/S^{(t_1)}_{\ell=0}$.
In comparison with the experiment, the theoretical models predict the
correct shape of the DCS. They underestimate the data around
$\vartheta_{\pmb{k}}=90^\circ$ and $\vartheta_{\pmb{k}}=270^\circ$,
however agree well at $\vartheta_{\pmb{k}}=0^\circ$ and
$\vartheta_{\pmb{k}}=180^\circ$ (along the polarization vectors). The
same observations apply for the DCS of the second ionization pathway
$|t_2|^2$ in panel \ref{fig4}(b). Figure \ref{fig4}(c) shows the
photoionization probability $|t_1+t_2|^2$ when both laser fields are
set to resonance with the respective $5s\rightarrow np$ transitions.
Here, the infrared field is again detuned by a small
$\delta\omega_{\rm IR}$ so that both amplitudes $t_1$ and $t_2$
have the same magnitude. In addition, we show here the result for
rather short laser pulses with 12 optical cycles.
\begin{figure}[t]
\includegraphics[width=1.0\columnwidth]{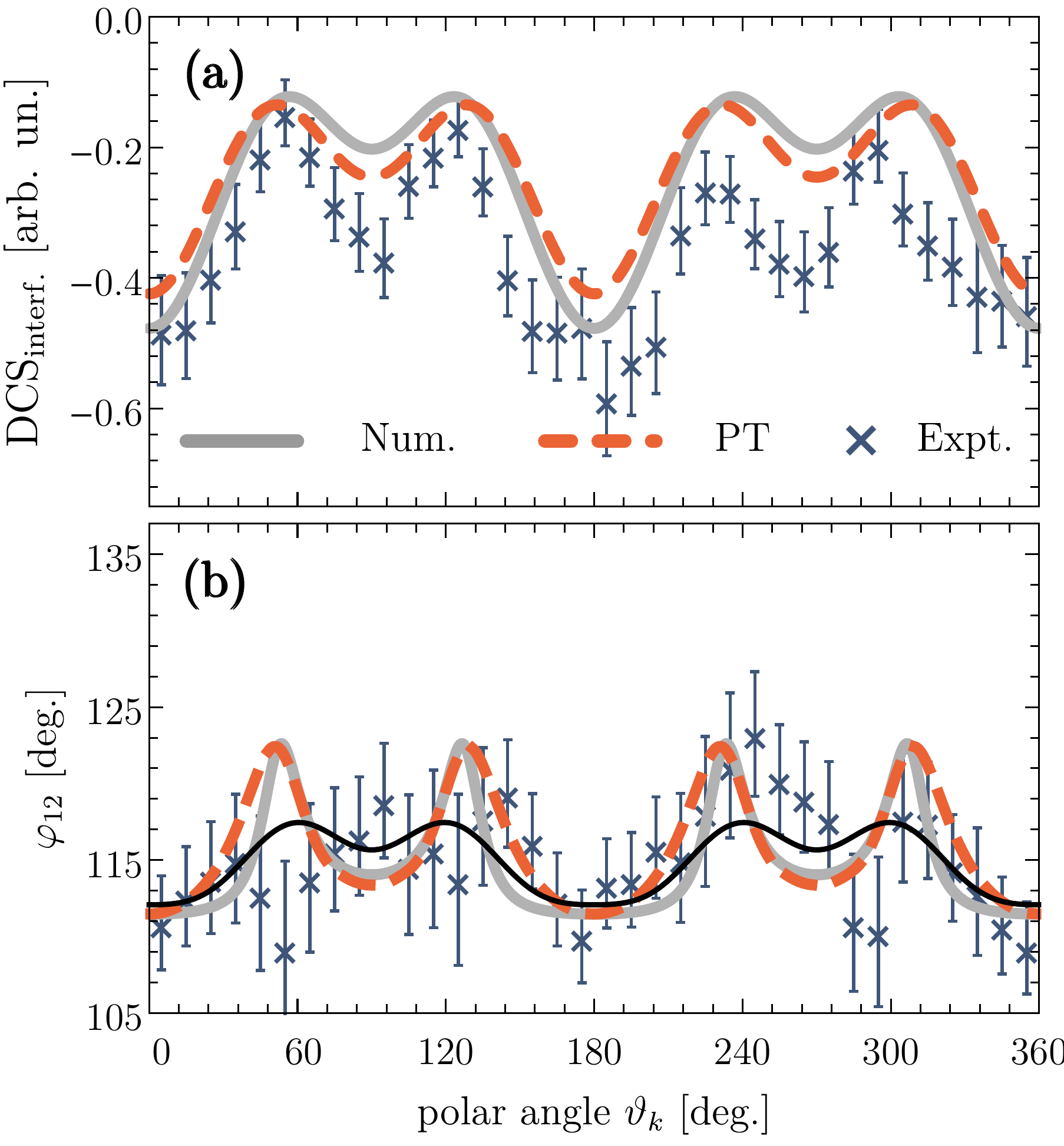}
\caption{(a) Interference term ${\rm DCS}_{\rm interf.}$ for  two
theoretical models and experimental data. Parameters are the same as
in Fig.\,\ref{fig4}. (b) Phase difference $\varphi_{12}$ between both
ionization pathway amplitudes $t_1(\vartheta_{\pmb{k}})$ and
$t_2(\vartheta_{\pmb{k}})$. The yellow line presents a fit of the
experimental data to the symmetry-adapted function
$f(\vartheta_{\pmb{k}})=\sum_{n=0}^2
a_{2n}\cos^{2n}(\vartheta_{\pmb{k}})$. The data points are reproduced from Ref.\cite{pursehousedynamic}.}
\label{fig5}
\end{figure}
Surprisingly, the DCS extracted from the short pulse calculations
agree very well with the smaller maxima around
$\vartheta_{\pmb{k}}=90^\circ$ and $\vartheta_{\pmb{k}}=270^\circ$.
However, this rather accidental agreement must be considered within the
experimental uncertainties.

In Fig.\,\ref{fig5}(a) we present the interference term ${\rm
DCS}_{\rm interf.}$ for the two developed theoretical models and the
experimental data. We see that the amplitude of the interference term
is clearly non-zero and varies from $13\%$ to $55\%$ of the
normalized signal shown in Fig.\,\ref{fig4}(c). In panel
Fig.\,\ref{fig5}(b) we present the corresponding phase difference
$\varphi_{12}(\vartheta_{\pmb{k}})$ between the two-photon transition
pathways $t_1$ and $t_2$. In comparison to all amplitudes, the
agreement is less satisfactory which can be explained by the relatively large
uncertainties due to error propagation through the arccos function
\cite{pursehousedynamic}. Interestingly the average value of the phase
shift is accurately reproduced by both calculations. Under these conditions the predicted angular variation is not very
pronounced and ranges from $110^\circ$ to $122^\circ$. Surprisingly,
 the models do not agree as well as for the DCS in
Fig.\,\ref{fig4}, which points to the extreme sensitivity of the
quantum phase to small changes in the transition matrix elements. The yellow curve represents a fit of the experimentally
obtained phase difference to the symmetry-adapted function $\sum_{n=0}^2 a_{2n}\cos^{2n}(\vartheta_{\pmb{k}})$.
It highlights the agreement with the theoretical prediction with respect to the general shape.

In the next section we will present mechanisms to manipulate, decrease and
increase this modulation.

\section{Manipulation of the quantum interference}
\subsection{Role of the energy gap}
\label{sec:5}
\begin{figure}[t]
\includegraphics[width=1.0\columnwidth]{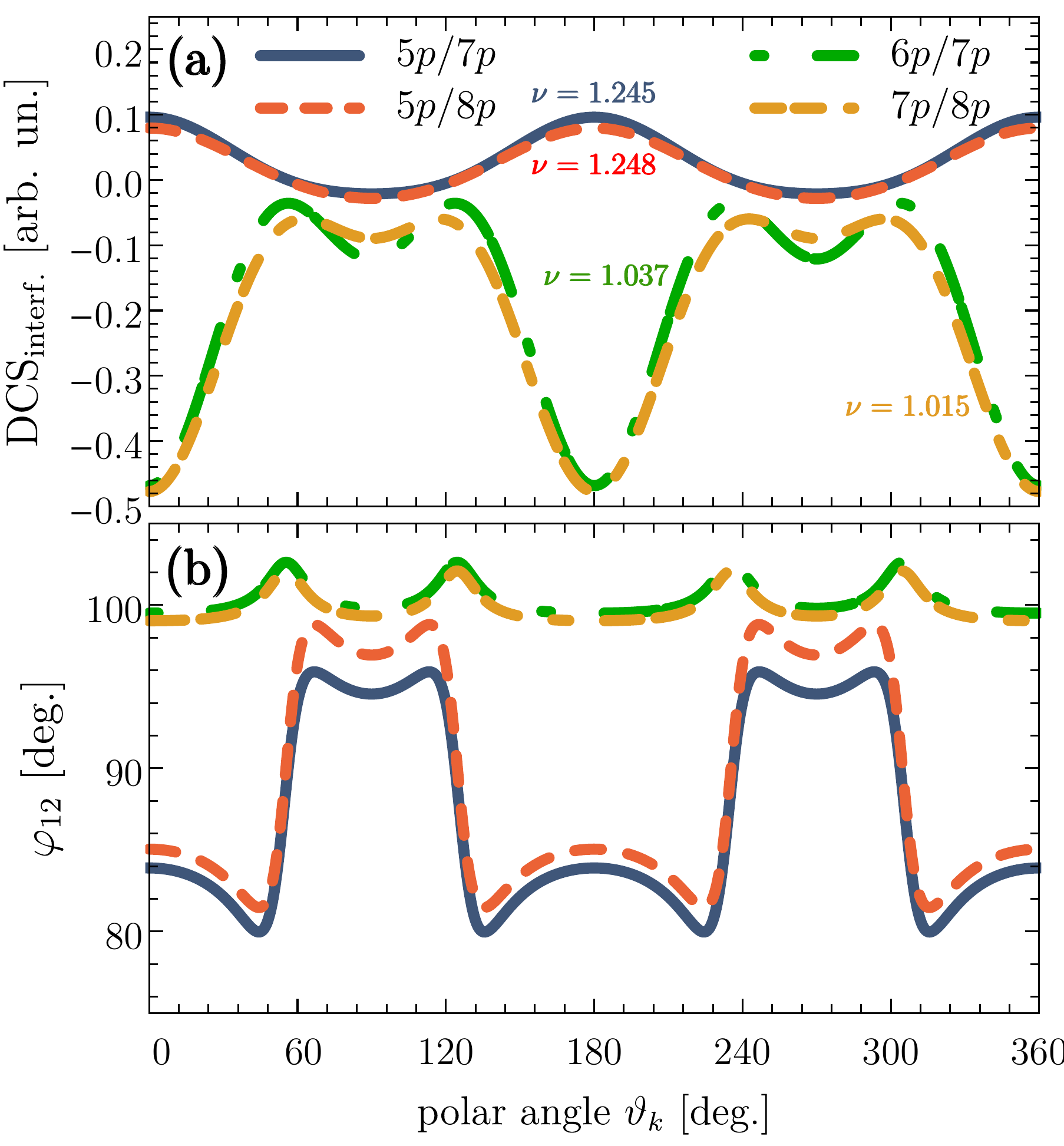}
\caption{Interference cross section (a) and phase difference (b)
$\varphi_{12}(\vartheta_{\pmb{k}})$ for different intermediate state
pairs extracted from the two-photon matrix element in
Eq.\,\eqref{eq:2PhotonMatrix}. The IR field is detuned to reach
equipollent pathway strengths.
The numbers in panel (a) represent the ratio
$\nu=(d^{(1)}_2/d^{(1)}_0):(d^{(2)}_2/d^{(2)}_0)$.}
\label{fig6}
\end{figure}
\begin{figure*}[t]
\includegraphics[width=0.95\textwidth]{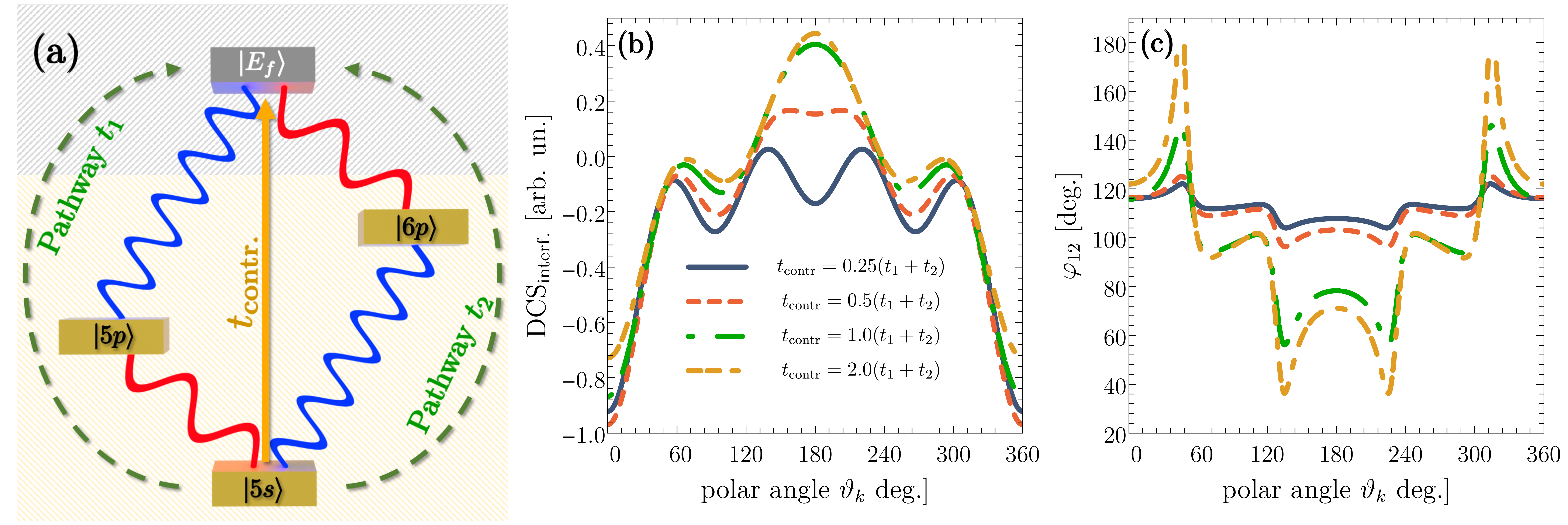}
\caption{(a) Addition of a third laser pulse (orange) with amplitude
$\mathcal{E}_3$ creates an additional third (direct) ionization pathway $t_{\rm
contr}$ to the final energy $E_{f}$. The blue and IR pulses are the same as
in Fig.\,\ref{fig0}. (b) Interference term ${\rm DCS}_{\rm interf}$
obtained from Eq.\,\eqref{eq:interference_mod} for different
strengths of the amplitude $t_{\rm contr}$ relative to
$t_1+t_2$. (c) Angular dependence of the associated interference
phase $\varphi_{12}$ between $t_1$ and $t_2$. All results are
obtained by numerically calculation with pulse lengths of 75 optical
cycles while the infrared and blue laser fields have the same field
amplitudes and detunings as in the previous sections.}
\label{fig7}
\end{figure*}
In Fig.\,\ref{fig6} we present interference studies on different
intermediate state pairs defining the two-color ionization amplitudes
$t_1$ and $t_2$. The blue and the red curves present the interference
between $5p/7p$ ($\Delta E= 1.88$\,eV) and $5p/8p$ ($\Delta E=
2.13$\,eV) states which have an increasing energy gap $\Delta E$ between them. The quantity $\nu=(d^{(t_1)}_{\ell=2}/d^{(t_1)}_{\ell=0})/
(d^{(t_2)}_{\ell=2}/d^{(t_2)}_{\ell=0})$ represent the ratios of the
bound-continuum reduced radial matrix elements between both pathways.
As expected for higher Rydberg states this ratio converges quickly to
the same number, i.e. the coupling of the $7p$ and $8p$ state to the
continuum is similar. The interference for both state pairs is
hence comparable. Note that the final energy
$E_f=E_{5s}+\hbar\omega_1 + \hbar\omega_2$ is larger in comparison to
the $5p/6p$ case (0.84\,eV and 1.09\,eV). Interestingly, ${\rm
DCS}_{\rm interf.}$ has a different shape and changes its sign in
both cases, which has an impact on the angular modulation of the phase
difference $\varphi_{12}$. Surprisingly while the amplitude is
smaller in comparison with the $5p/6p$ case shown in
Fig.\,\ref{fig5}, the angular variation of the phase is significantly more
pronounced, ranging from $80^\circ$ to $100^\circ$. This highlights
the change of sign of the interference term (negative sign of the
argument of the arccos function means a phase larger than
$90^\circ$).

The green and orange curves represent cases when we decrease the
energy gap. We chose the intermediate state pairs $6p/7p$ ($\Delta
E=0.51$\,eV, $E_f=2.22$\,eV) and $7p/8p$ ($\Delta E=0.25$\,eV,
$E_f=2.97$\,eV). Intriguingly, the interference cross section remain
unaffected when changing to higher lying state pairs. The amplitudes
of ${\rm DCS}_{\rm interf.}$ ranges in both cases from 5\% to
50\% and is negative. However, the angular modulation of the phase
difference $\varphi_{12}$ decreases drastically. We address this
development with the dipole matrix element ratio $\nu$ which
converges rapidly to 1, meaning the ratio between the $\ell=0$
and $\ell=2$ angular channels is  nearly the same for both
ionization pathways $t_1$ and $t_2$. According to
Eq.\,\eqref{eq:phase} the individual phase shapes are then
comparable.

From these results we learn that the angular modulation of the phase
difference depends critically on the quantity $\nu$, while the overall
amplitude of the interference term (${\rm DCS}_{\rm interf.}$) is
more robust and reveals a dependence on the energy gap between the
intermediate state pair defining $t_1$ and $t_2$.

\subsection{laser-driven perturbation of ionization pathways}

Another method for dynamic control of the interference phenomena is
the addition of a third control pulse. As represented in the scheme
in Fig.\,\ref{fig7}(a) the corresponding parameters are chosen in a way
that  initiates a weak one-photon process directly into the
continuum, so that $\hbar\omega_{\rm contr.}=(E_f-E_{5s})/\hbar$.
The field amplitude $\mathcal{E}_{\rm contr}$ is hence chosen in a
way that the corresponding transition amplitude $t_{\rm contr}$ is of
the same magnitude as that of the two-color pathways $t_1$
and $t_2$. As indicated in the modified scheme in
Fig.\,\ref{fig7}(a), in the presence of all fields the total
amplitude is given by the coherent sum $t_{\rm all}=t_1+t_2+t_{\rm
contr.}$. To access the desired interference term ${\rm
DCS}_{\rm interf}=t_1t_2^* + t_1^*t_2$ one has to then perform four
different measurements: (i) with all fields on resonance to the
intermediate and final states respectively, (ii) with the blue
field to the $6p$ state off-resonance, (iii) with the red field
to the $5p$ state off-resonance, and (iv) with both red and blue
fields off-resonant (thereby blocking both
pathways $t_1$ and $t_2$). The interference term is then given by
\begin{equation}
{\rm DCS}_{\rm interf}=|t_{\rm all}|^2 - (|t_{\rm (ii)}|^2 + |t_{\rm
(iii)}|^2 - |t_{\rm (iv)}|^2)
\label{eq:interference_mod}
\end{equation}
with $t_{\rm (ii)}=t_1 + t_{\rm contr}$, $t_{\rm (iii)}=t_2 + t_{\rm
contr}$ and $t_{\rm (iv)}=t_{\rm contr}$.

In Fig.\,\ref{fig7}(b) we present the interference term for
different strengths of the perturbation by the third (control)
laser field. As expected, the additional one-photon ionization route
has a large impact on the quantum interference between $t_1$ and $t_2$.
Here, the field amplitude $\mathcal{E}_{\rm contr}$ has to be very
low so that the associated $t_{\rm contr}$ is of the same magnitude
as the two-photon pathways $t_1$ and $t_2$. In comparison to the
unperturbed case shown in Fig.\ref{fig5}(a), the magnitude of the
interference is increased in the presence of the control field. In
strong contrast to the previous findings, even the sign of the ${\rm
DCS}_{\rm interf}$ can be changed by the effect of the additional
one-photon process when the field amplitude is sufficiently large.
It is therefore not surprising that the large impact seen here is directly
transferred to the phase $\varphi_{12}$ associated with the quantum
interference. As shown in Fig.\,\ref{fig7}(c), the angular variation is
drastically increased due to the action of the controlling field. In the
original experiment and theoretical treatment the modulation in the
polar angle was smaller than 20$^\circ$. Now we obtain strongly
pronounced phase peaks and a rather complex angular structure of
$\varphi_{12}$ with a variation covering more than 140$^\circ$.

\begin{figure}[t]
\includegraphics[width=0.95\columnwidth]{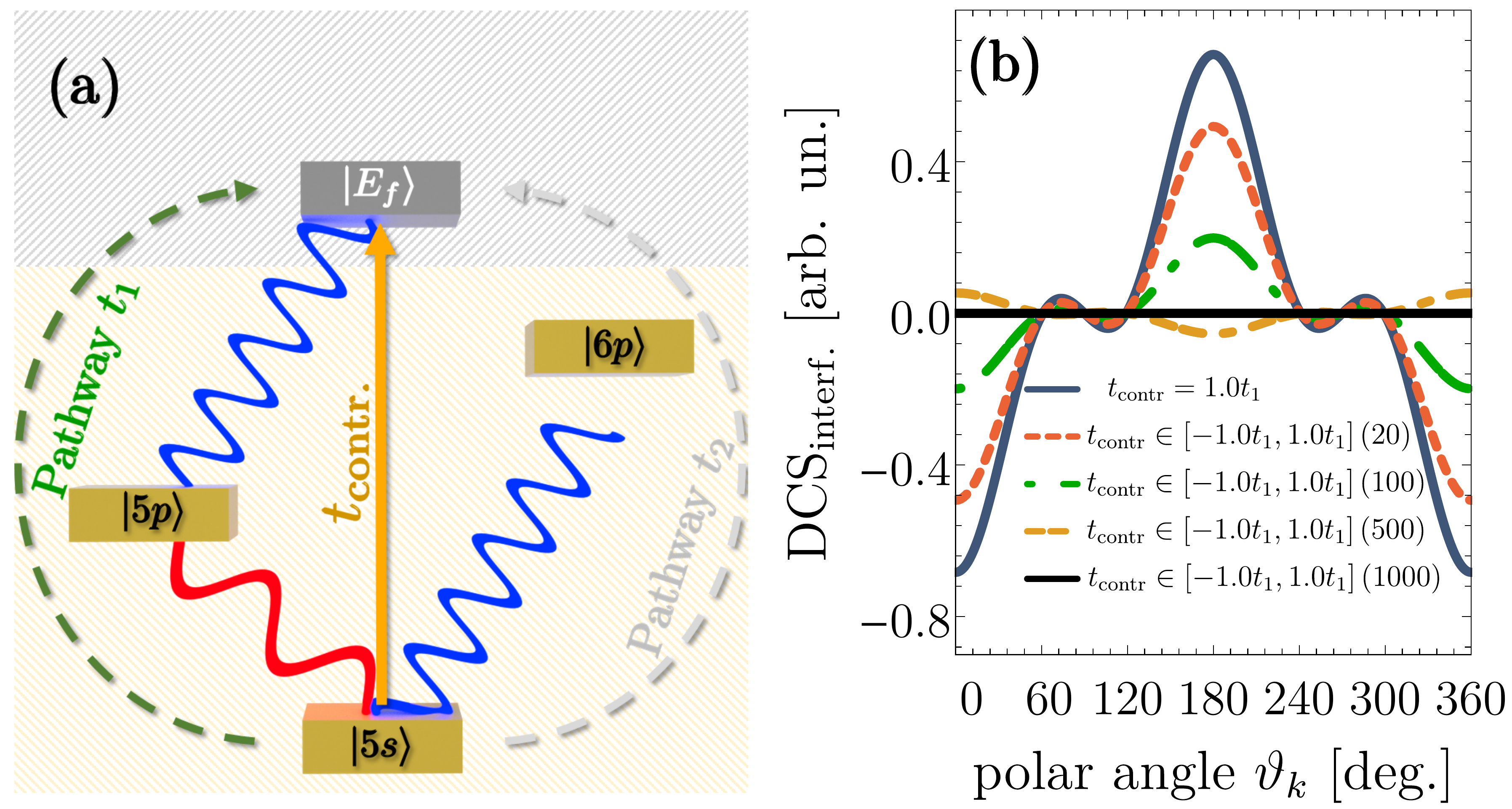}
\caption{Stochastic nature of the interference effect. In this example one of
the ionization pathways is closed by slight detuning of the laser
field. We hence
find interference between the two-photon amplitude (pathway 1) and
the one-photon transition initiated by the third control field
$E_3(t)$ (scheme panel a). (b) The stochastic nature of the interference
effect: the amplitude of the one-photon process is randomly varied between
$-1.0t_1$ and $+1.0t_1$. The curves show the dependence on the number
of random amplitudes with the reference to $t_{\rm contr.}=1.0t_1$
(blue line). Numbers in the round brackets denote the number of
amplitudes used for the statistical average (see text for details).}
\label{fig8}
\end{figure}

The addition of the third laser field helps to emphasize that the
interference effect is not robust to statistical fluctuations, but is
unique for every set of laser parameters.
For this purpose we slightly detuned the blue laser field so that
the $6p$ state was not excited [cf.\,Fig\,\ref{fig8}(a)]. The
resulting interference phenomenon in the ionization channel then stems
from the superposition of the two-photon pathway 1 (via $5p$
photoexcitation) and the one-photon direct photoionization amplitude
mediated by $E_{\rm contr.}$. By tuning the parameters of the third field so
that the transition strength of the amplitude $t_{\rm contr.}$ is
equal to $t_1$ we obtain a characteristic interference as shown by
the dark blue curve in Fig.\,\ref{fig8}(b). We then varied the
amplitude of the control field in a way that $t_{\rm
contr.}\in[-1.0t_1,1.0t_1]$ by use of a random number generator. The
additional curves in the figure show a statistical average based on the total number of
random amplitudes input to the model. One can clearly see a trend that increasing the
number of random events decreases the interference effect. This shows
that for an infinite number of measurements the resulting
interference would disappear.

\section{Conclusions}

In this paper we have presented a theoretical investigation of
experimental interference studies in a single Rubidium atom.
We have systematically demonstrated the transition from the short-pulse into
the continuous wave regime, and the evolution of the occupation numbers,
photoionization probability and quantum phase under an increase of
the pulse lengths. We find that for pulse lengths of more than ten
optical cycles the theoretical description of the ionization scheme
via the two-photon matrix element in the frequency regime is
sufficient, and that this provides all the physical information
required for interference studies.

Our theoretical model provides generally good agreement with the
experimental data and predicts a pronounced interference amplitude
${\rm DCS}_{\rm interf.}$ while the angular variation of the
associated phase difference $\varphi_{12}$ is relatively weak. In this
treatment we have developed various strategies to manipulate the quantum
interference between both photoionization pathways $t_1$ and $t_2$.
As an example, we can change the populations of the intermediate
states by laser detuning which introduces an imbalance between both
pathways so as to change the interference phenomena. Further,
by choosing different state pairs $n_1p$ and $n_2p$ we change the
energy difference and coupling to the continuum, which again markedly
changes the quantum interference. A new method which does not
change the intermediate state pairs and the parameters of the blue
and infrared laser fields is the addition of a third control laser
field which perturbs the original transition pathways $t_1$ and $t_2$. An
appropriate tuning of the corresponding one-photon transition
amplitude into the continuum can even invert the sign of the
interference amplitude, as well as produce a much more pronounced
angular variation of the interference phase. There is no analogy to
this third control in the conventional double slit experiments.

As well as the addition of a third pulse, there are several other
possibilities to explore the behaviour of the interference phenomenon.
As an example, one can show that in Alkali atoms quadrupole
transitions into the continuum reveal Cooper minima at kinetic
energies below 1eV. Thus, one can choose intermediate state pairs
which lead to final energies in the region of such Cooper minima
while such quadrupole transitions at low intensity are generally
accessible by structured light fields \cite{schmiegelow2016transfer}.
Further work will hence be dedicated to studying these two-pathway interference
effects with inhomogeneous light-induced quadrupole transitions.

\section*{Acknowledgements}

This work was partially supported by the DFG through SPP 1840 and SFB TRR 227. The EPSRC
U.K. is acknowledged for current funding through Grant No. R120272.


%

\end{document}